# Oxidation-State Dynamics and Emerging Patterns in Magnetite

Emre Gürsoy, Gregor B. Vonbun-Feldbauer, and Robert H. Meißner*

ACCESS | Metrics & More | Article Recommendations | Supporting Information

**ABSTRACT:** Magnetite is an important mineral with many interesting applications related to its magnetic, electrical, and thermal properties. Typically studied by electronic structure calculations, these methods are unable to capture the complex ion dynamics at relevant temperatures, time, and length scales. We present a hybrid Monte Carlo/molecular dynamics (MC/MD) method based on iron oxidation-state swapping for accurate atomistic modeling of bulk magnetite, magnetite surfaces, and nanoparticles that captures the complex ionic dynamics. By comparing the oxidation-state patterns with those obtained from density functional theory, we confirmed the accuracy of our approach. Lattice distortions leading to the stabilization of excess charges and a critical surface thickness at which the oxidation states transition from ordered to disordered were observed. This simple yet efficient approach paves the way for elucidating aspects of oxidation-state ordering of inverse spinel structures in general and battery materials in particular.

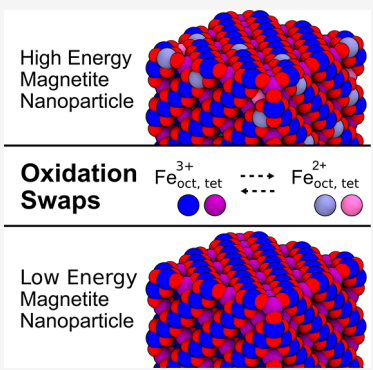

Oxidation Swaps  $Fe^{3+}_{oct, tet}$  $Fe^{2+}_{oct, tet}$

High Energy Magnetite Nanoparticle

Low Energy Magnetite Nanoparticle

Historically, magnetite is probably the first known magnetic material and has been studied intensively since the pioneering work of Bragg.[1] It has become an indispensable biocompatible material useful in many biomedical applications, from theranostics or drug delivery agents[2] to magnetic resonance imaging (MRI).[3] It also has great potential for heterogeneous catalysis[4,5] and is used as a building block in hierarchical nanocomposites. Such materials achieve exceptionally good mechanical properties,[6−8] and by choosing a suitable nanoparticle morphology, the final hierarchical structure of the nanocomposite could be modified.[9,10] Magnetite nanoparticles are also useful for water purification, for example to remove glyphosate from water.[11]

At room temperature, magnetite is a mixed valence metal oxide with an inverse spinel structure $Fe^{3+}_{tet}[Fe^{3+}Fe^{2+}]_{oct}O_4$. While one-third of the irons are tetrahedrally coordinated exhibiting $Fe^{3+}$, two-thirds of the irons are octahedrally coordinated showing a mixture of $Fe^{2+}$ and $Fe^{3+}$ at a ratio of 1:1. At around 125 K, magnetite undergoes the Verwey transition, which is associated with changes in its magnetic, electrical, and thermal properties.[12] This transition is strongly dependent on the stoichiometric composition of the homogeneous spinel phase[13] and accompanied by an increase in electrical conductivity of up to 2 orders of magnitude, i.e., from an insulator to a semimetal. It is generally accepted that the electrical conductivity of magnetite is related to the exchange of valence electrons between $Fe^{3+}$ and $Fe^{2+}$ on the octahedral sublattice.[14] More recently, even a thermal transfer of electrons from octahedral to tetrahedral iron sites has been confirmed by X-ray spectroscopy.[15]

In addition to the complex bulk phenomena of magnetite, its surfaces exhibit particularly interesting features, as revealed by recent experiments. The two most important magnetite surfaces are presumably the (001) and (111) surfaces. For the (001) surface, two surface structure models are commonly used, the distorted bulk truncation (DBT) model[16] and the subsurface cation vacancy (SCV) reconstruction.[17] While the SCV structure is usually found for clean surfaces under ultra-high-vacuum conditions, a DBT can be stabilized by adsorbates.[18−20] For the (111) surfaces, the stable termination depends strongly on the environment and the preparation conditions.[21−26] The $Fe_{tet1}$ termination is found for a wide range of relevant oxygen pressures.[24]

The complex relationship between charge ordering and valence electron interactions between $Fe^{3+}$ and $Fe^{2+}$ in magnetite has only recently been explored by quantum chemical calculations,[27−30] and with a few exceptions, little progress has been made in developing accurate force fields.[23,30,31] These quantum chemical calculations agree well with the experimental results, but it is not practical to apply these methods to more realistic systems containing thousands of atoms. Methods that can handle thousands of atoms while balancing accuracy and efficiency are, therefore, in high demand.

We present a simple yet effective approach that combines Monte Carlo (MC) with (i) simulated annealing to obtain energetically minimized oxidation-state configurations and with (ii) molecular dynamics (MD) to study dynamically emerging oxidation-state patterns. In this approach, partial atomic charges of Fe are swapped according to the probability of accepting a swap, $P_{acc}$, given by the underlying Metropolis-Hastings algorithm:

$$P_{acc}(q_{Fe}^{2+} \leftrightarrow q_{Fe}^{3+}) = \min\left\{1, \exp\left(-\frac{U_f - U_i}{k_B T^{MC}}\right)\right\} \qquad (1)$$

which implicitly changes thus their oxidation state and overcomes an inherent limitation of empirical force fields, namely that the partial charges are "fixed". $k_B$ is the Boltzmann constant, and $U_i$ and $U_f$ are the potential energies given by the force field before and after a swap, respectively. By choosing a sufficiently high-temperature $T^{MC}$ in eq 1 combined with a sophisticated annealing procedure, it is possible to find a reasonable minimized configuration or to sample the configurational phase space when combined in a hybrid MC/MD approach, in which $n_{swaps}$ oxidation swaps are performed before each MD step. Methodologically, this is implemented using the `fix atom/swap`[32] available in LAMMPS (see SI for an example script). It is important to note that the full energy of the system must be recalculated for each swap, since Coulombic interactions are inherently long-range and cannot be easily localized to a subset of atoms. This opens the door to studying oxidation-state patterns arising from the ability of Fe ions to adapt to their electrostatic environment by what we denote here as oxidation-state swapping.

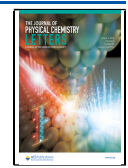

For the description of magnetite, we rely partly on parameters already available in ClayFF.[33,34] Of particular importance for the approach presented here are the partial charges derived in ClayFF. However, ClayFF has not been parametrized specifically for magnetite, and thus, among others, no partial atomic charge is available for $Fe^{2+}$. As will be shown below, it turns out to be a reasonable choice to choose the opposite partial atomic charge of $O^{2-}$ for $Fe^{2+}$, since this, perhaps most importantly, ensures a charge-neutral bulk magnetite unit cell. Furthermore, using this partial atomic charge for $Fe^{2+}$ results in a good description of bulk magnetite, magnetite surfaces, and nanoparticles. The complex energy landscape associated with the almost infinite possibilities of oxidation-state configurations in magnetite at a finite temperature above the Verwey transition is hence explored by using a hybrid MD/MC approach, in which $n_{swaps}$ oxidation swaps are performed before each MD step. Unless otherwise stated, MD simulations are performed in the NVT ensemble, which allows atoms to move according to the given temperature $T$ but with a fixed volume $V$ of the simulation cell and a constant number of atoms $N$, by solving Newton's equations of motion numerically using the velocity-Verlet algorithm. However, if the minimized oxidation-state configuration is sought, e.g., for comparison with density functional theory (DFT), this is achieved by fixing the atoms (typically at their ideal lattice positions) and gradually lowering the MC temperature $T^{MC}$ using an exponential annealing scheme (see SI for more details).[35]

An important assumption we make is that charge neutrality is maintained, even when magnetite is not stoichiometric. While this is a common observation for magnetite surfaces in experiments, it is even more prominent in simulations. This is mainly due to the limitations of the system size and the associated problem of nowhere to place the charge-balancing ions, which are most likely to be found at defects or grain boundaries somewhere in the bulk.[36] In addition, artifacts due to nonconverging electrostatics are avoided, as a charge-neutral system is usually required in atomistic simulations with force fields when using periodic boundary conditions along with a long-range solver for electrostatics. Furthermore, in contrast to the common approach of defining a formal oxidation state of $Fe^{2.5+}$ for the octahedral irons, more realistic explicit partial atomic charges are used for $Fe^{2+}$ and $Fe^{3+}$. Consequently, we use the following relation to ensure charge neutrality:

$$n_{Fe}^{2+} q^{2+} + n_{Fe}^{3+} q^{3+} + n_O q^{2-} = 0 \qquad (2)$$

where $q^{3+}$, $q^{2+}$, and $q^{2-}$ are the partial atomic charges representing the oxidation states of iron and oxygen as derived from DFT calculations.[30] $n_{Fe}^{2+}$, $n_{Fe}^{3+}$, and $n_O$ are the total numbers of iron atoms, with their oxidation state indicated by the superscript, and oxygen atoms in the simulation cell. We assume that atoms of the same oxidation state have the same partial atomic charge. The Bader charges of $Fe^{2+}$ and $Fe^{3+}$ differ significantly. In contrast, the variations of the Bader charges within $Fe^{2+}$ or $Fe^{3+}$ are small and depend on the chemical environment. Nevertheless, it is reasonable to neglect these small charge differences (see SI for more details) because the Bader charges do not overlap, keeping the approach efficient, simple, and versatile. In addition, we assume here that the partial atomic charge of oxygen is the same as that of $Fe^{2+}$, but with the opposite sign $q^{2-} \equiv -q^{2+}$. Using $n_{Fe}^{2+} + n_{Fe}^{3+} = n_{Fe}$ in eq 2, a rather general relation of the ratio of $Fe^{2+}$ and $Fe^{3+}$ for nonstoichiometric magnetite, depending on the amount of iron and oxygen in the compound, is obtained:

$$\frac{n_{Fe}^{2+}}{n_{Fe}^{3+}} = \frac{q^{3+} n_{Fe} - q^{2+} n_O}{q^{2+}(n_O - n_{Fe})} \qquad (3)$$

If not otherwise noted, initial magnetite structures are generated by selecting a random set from all iron atoms based on the ratio in eq 3 and subsequently assigning them a partial point charge of $q^{3+}$; the remaining irons are set to $q^{2+}$.

We begin the discussion of the results by applying this framework to a bulk magnetite system. A bulk magnetite simulation cell is generated by replicating the $Fd\bar{3}m$ unit cell three times in each dimension. Starting from a random (but stoichiometric) distribution of oxidation states on all iron sites and fixing the atoms to their ideal lattice positions, a minimized oxidation-state configuration was obtained using an exponential simulated annealing scheme starting from $T_0^{MC} = 10^5$ K. Around a critical $T_c^{MC}$ of 459 K, the oxidation swap acceptance rate dropped to zero, indicating that a minimized oxidation-state configuration is reached (cf. Figure S2). As expected, the minimized oxidation-state configuration has a 1:1 ratio of $Fe^{2+}$ and $Fe^{3+}$ on the octahedral sites, and all tetrahedral irons are $Fe^{3+}$. We observed some dependency of this critical temperature on the system size (cf. Figure S2a). However, it is difficult to distinguish whether this is due to different MC settings affecting convergence or the finite size of the systems. Determining an exact value would require larger simulations with more swaps, which were not feasible within the scope of this work and, moreover, would not change the main message of the more general insights presented here. Interestingly, when the system is reheated from its minimized and relaxed configuration, i.e., the atoms were allowed to move during the relaxation after minimizing the oxidation states but not during the reheating, an increase in the onset temperature of the oxidation swaps is observed (cf. Figure S2b). We hypothesize that excess charges, i.e., either a $Fe^{2+}$ or a $Fe^{3+}$, are stabilized by local lattice deformations around them and delay the onset of the swaps. Such local lattice deformations stabilizing an excess charge are commonly referred to as small polarons[37]—hence we refer to this phenomenon as "small polaron-like" behavior in our simulations. Unfortunately, our force field is rather limited in its interpretation of polarons and can probably only address the electrostatic effect of them, neglecting the spin, orbital, and magnetic effects. It is an interesting topic in itself[38] and even occurs in natural magnetite[39] but has too many implications to be covered comprehensively with this approach, e.g., the coupling of such lattice deformations to phonons,[40] electronic collective modes,[41] or magnetism.[42,43]

To confirm that local lattice deformations stabilize the excess charge, two strategies were followed. In both cases, starting from the minimized oxidation-state configuration, the atomic positions of the magnetite ions were allowed to move using the hybrid MC/MD approach. While the MD temperature was kept at 300 K, the MC temperature was kept either below the critical temperature $T^{MC} = 300\ K < T_c^{MC}$ or above $T^{MC} = 10^5\ K \gg T_c^{MC}$. Note that in this case we restrict swaps to octahedral sites only, and a rather strong coupling with the thermostat is required. The resulting radial distribution functions (RDFs) of $Fe^{2+}$ and $Fe^{3+}$ have been calculated and compared with the ideal crystal in Figure 1. For $T^{MC} < T_c^{MC}$, the average distance between neighboring $Fe^{2+}$ ions becomes shorter by about 0.05 Å (cf. Figure 1a), and the distance between $Fe^{3+}$ ions becomes slightly larger compared to an ideal lattice.[44] This is in good agreement with the results of Senn et al.,[44] who also observed some anomalous Fe—Fe distance shortening and lengthening of mainly $Fe^{3+}$—$Fe^{3+}$ distances associated with polaron formation. On the contrary, when using $T^{MC} \gg T_c^{MC}$, the anomalous shortening of the $Fe^{2+}$—$Fe^{2+}$ distances and lengthening of $Fe^{3+}$—$Fe^{3+}$ disappear, as illustrated in Figure 1b.

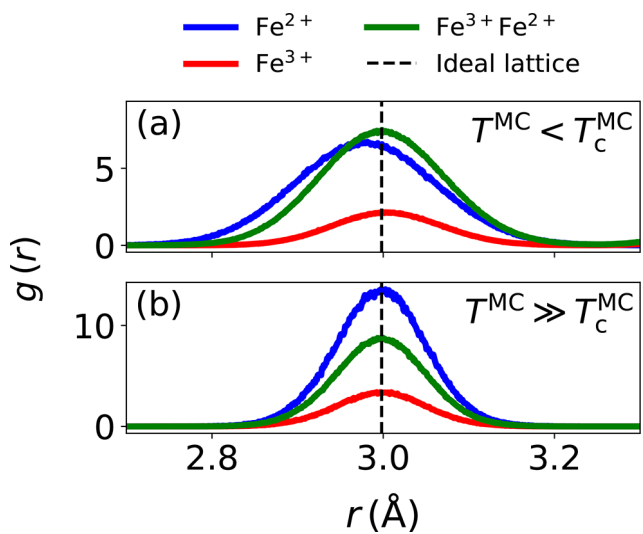

**Figure 1.** First peaks of bulk magnetite RDFs, $g(r)$, between the iron species indicated in the legend. (a) RDF obtained from a hybrid MD/MC simulation starting from an initially minimized oxidation-state configuration and keeping the MC temperature $T^{MC}$ below the critical temperature $T_c^{MC}$ at an elevated MC temperature of $10^5$ K above the critical MC temperature while maintaining 300 K in the MD in both cases. In addition, the first RDF peak of an ideal (i.e., undistorted) lattice is shown by a dashed black line. Full RDFs are shown in Figure S1.

Allowing for swaps between all iron sites, similar to what was observed by Elnaggar et al.,[15] tetrahedral iron sites are increasingly occupied by $Fe^{2+}$ with increasing temperature. In particular, the amount of $Fe_{tet}^{2+}$ depends on $T^{MC}$ and follows the relationship given in ref 43: $(Fe_{1-x}^{3+}Fe_x^{2+})_{tet}[Fe_{1+x}^{3+}Fe_{1-x}^{2+}]_{oct}O_4$. A $T^{MC}$ of 4000 K corresponds to the experimental temperature range between 330 and 880 K (called Regime II in ref 15) with a value of $x = 0.125$ at 840 K.[43] Note that $T^{MC}$ here is typically an order of magnitude larger than the experimentally observed temperatures.

In summary, the hybrid MC/MD approach appears to be able to mimic the distribution of localized electrons in high-temperature cubic magnetite over linear three-Fe units called "trimerons" and can reproduce the resulting anomalous shortening of some Fe—Fe distances and lengthening of others observed not only at temperatures below the Verwey temperature[44] but also above it.[45] Interestingly, while this effect is explained by a rather complex population of orbitals in magnetite, i.e., Jahn—Teller distortions,[46] it is surprising that our relatively simple force field approach is able to partially capture it. It seems that the inverse spinel lattice of magnetite is somewhat peculiar and allows such specific oxidation-state ordering purely due to electrostatic effects. As a consequence, below a critical MC temperature, polarons are stabilized even though the atoms are able to move. In addition, at this temperature, oxidation swaps can no longer occur. $T_c^{MC}$ is obviously not the Verwey transition temperature in this case. Although it is most likely related to the Verwey transition, due to inaccuracies in the force field, neglect of other important quantum mechanical effects (e.g., orbital and magnetic effects), and probably also finite size effects, our approach is certainly not able to fully account for it but allows some interesting insights into the electrostatic effects on small polaron formation.

With this in mind, we have tried to apply this approach to the $Cc$ magnetite structure below the Verwey temperature, but we have neither observed the monoclinicity nor the layered structure of $Fe^{2+}$ and $Fe^{3+}$ on the octahedral sites as observed in Senn et al.[44] This seems to be the limiting case where our approach considering only oxidation-state ordering starts to fail and orbital ordering[47] or magnetic effects[42] in magnetite become the more dominant effect in trimeron ordering[43]—although this has been discussed controversially before.

Magnetite surfaces are inherently nonstoichiometric due to the $Fe^{2+}/Fe^{3+}$ imbalance with respect to oxygen.[17,21,36] Consequently, in common force field simulation approaches, surface modification is associated with a nonstoichiometric composition of the resulting simulation system, which as a consequence is typically not charge neutral, as one is usually unable to add or remove the necessary neutralizing charge anywhere in the bulk due to limited system sizes. However, it should be noted that the Verwey temperature was found to be very sensitive to chemical composition, but the inverse spinel structure of magnetite tolerates small deviations from ideal stoichiometry.[48]

We validate the previously introduced charge-neutrality condition in eq 3 by comparing the resulting $n_{Fe}^{2+}/n_{Fe}^{3+}$ ratio with the ratio obtained from DFT calculations for the same magnetite surface in Figure 2a. The charge-neutrality equation gives almost identical $n_{Fe}^{2+}/n_{Fe}^{3+}$ ratios compared with the DFT results. A deviation from the DFT results is observed only for very thin (001)-SCV surfaces, where the charge-neutrality equation predicts a nonphysically negative $n_{Fe^{2+}}/n_{Fe^{3+}}$ value. Recently, such freestanding thin sheets of magnetite, known as magnetene, have attracted some attention.[49] Note that for such a thin (001)-SCV slab, DFT indicates that there is only one charge value, resembling more a $Fe^{3+}$ oxidation state, which consequently gives $n_{Fe^{2+}}/n_{Fe^{3+}} = 0$.

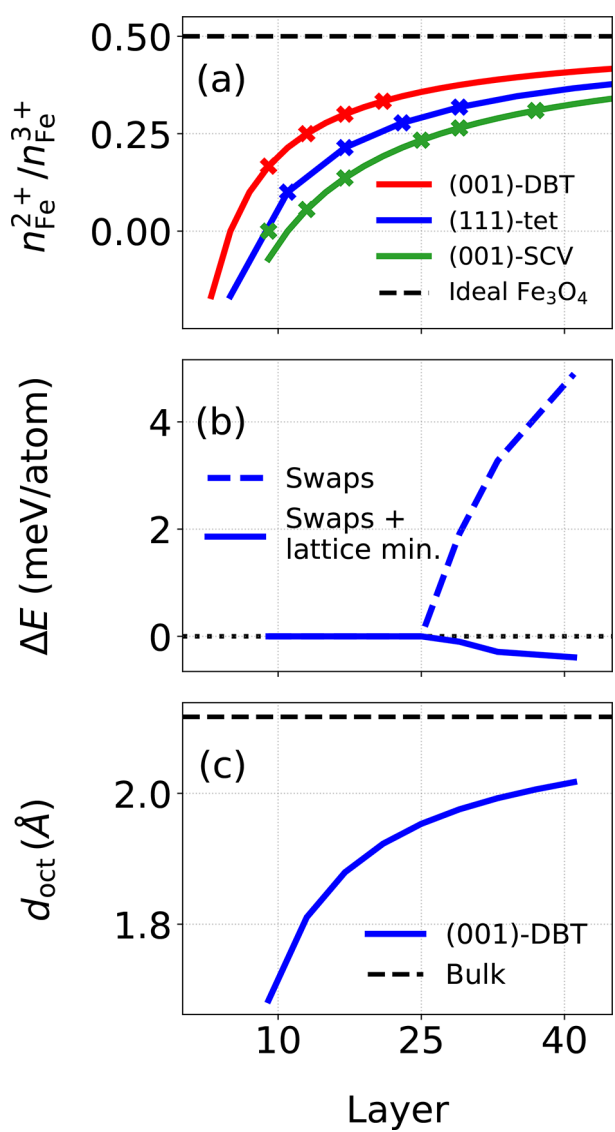

**Figure 2.** (a) Stoichiometry of magnetite surfaces. Ideal magnetite stoichiometry is indicated by the black dashed line. $n_{Fe^{2+}}/n_{Fe^{3+}}$ ratios derived from charge neutrality, given by eq 3, are represented by lines; ratios derived from DFT are represented by crosses. (b) Potential energy differences ($\Delta E$) between layered and minimized oxidation-state configurations of (001)-DBT surfaces of varying thickness. (c) Octahedral iron layer spacings ($d_{oct}$) of (001)-DBT surfaces as a function of the number of atomic layers. Layer spacing in bulk magnetite is indicated by the black dashed line.

Previously, Konuk et al.[30] studied the charge distribution of Fe ions on magnetite surfaces. We recalculated these systems with fewer computational constraints to avoid ambiguity in the assignment of charge values and included larger systems. See the Methods section of the Supporting Information for more details. Already, Konuk et al.[30] observed from their DFT calculations on (001)-DBT surfaces that the first two octahedral layers are entirely $Fe^{3+}$ in agreement with other literature.[17,20] The subsequent octahedral layers alternate between being fully $Fe^{3+}$ and $Fe^{2+}$. In contrast Liu and Di Valentin[20] also using DFT have observed configurations for a 17-layer slab with a bulk-like central region with fixed bulk layer distances where $Fe^{2+}$ and $Fe^{3+}$ ions are distributed well-ordered in a 1:1 ratio in each octahedral layer. The energy differences between oxidation-state configurations are likely to be quite small, and thus, the distributions are probably very sensitive to the computational settings and small differences in the atomic structure. In our current DFT calculations, we did not see a bulk-like region for a 17-layer slab neither when using bulk layer distances in the central region nor for relaxing the whole slab, but the first mixed layers appear from 21 layers on; however, a fully bulk-like central region is not visible yet. The aim of the following is to test whether the oxidation-state swapping method can determine a minimum layer thickness which defines the limit at which a slab still contains a bulk-like central region with mixed oxidation-state ordering, where $Fe^{2+}$ and $Fe^{3+}$ ions are randomly distributed in a 1:1 ratio in each octahedral layer. To achieve this, we first manually ordered the oxidation-state configurations of the (001)-DBT surfaces to resemble the oxidation-state ordering observed from DFT. Alternatively, simulated annealing, as previously used for bulk structures, was applied to equivalent surfaces with initially randomly distributed oxidation states and resulted in comparable minimized oxidation-state configurations. Configurations from both approaches are structurally minimized to ensure that the atomic positions of the magnetite atoms are adapted to their electrostatic environment. The resulting energy difference $\Delta E$ between the two approaches, i.e., a layered and a minimized oxidation-state configuration, is then compared and shown in Figure 2b. In (001)-DBT slabs larger than 25 atomic layers, a transition from the surface to bulk appears to form. When the layering of these structures is examined in more detail (cf. Figure S3), it is clear that for these larger slabs a more disordered central part of the slab becomes increasingly more favorable than a layered structure.

In the smaller (001) DBT slabs, the oxidation-state layering effectively maximizes the number of small polaron-like deformations by surrounding each $Fe^{2+}$ with two $Fe^{3+}$ in the layer below and above. This should also result in a smaller distance between the octahedral layers. Indeed, as indicated by $d_{oct} = L/n_{oct}$ in Figure 2c, where $L$ is the surface thickness and $n_{oct}$ is the number of octahedral layers, a decrease in the general layer spacing is observed for more layered (and thinner) slabs. This could be an interesting aspect to be experimentally verified for thin magnetite films. In DFT calculations with fixed central layers, this behavior is hindered.

To investigate the surface-to-bulk transition in a (001)-DBT surface in more detail, Figure 3 shows, for three (001)-DBT surfaces of different thicknesses, the ratio of $Fe_{oct}^{3+}$ ions in each octahedral layer, $n_{Fe_{oct}^{3+}}/n_{Fe_{oct}}$, of the respective surface. In contrast to the layered oxidation-state pattern of the 25L-(001)-DBT subsurface layers in Figure 3a, the 33L-(001)-DBT subsurface layers in Figure 3b are neither occupied completely by $Fe_{oct}^{2+}$ nor $Fe_{oct}^{3+}$. For the thicker 41L-(001)-DBT slab in Figure 3c, as one moves toward the center of the surface, the $n_{Fe_{oct}^{3+}}/n_{Fe_{oct}}$ ratio approaches 0.5 already after a few surface layers, which we then call a "bulk-like" oxidation state. Furthermore, in the 41L-(001)-DBT surface, some $Fe^{2+}$ ions are observed already in the second $Fe_{oct}$ layer. Note that the minimized oxidation-state configuration of the (001)-DBT-

$$n^{3+}_{Fe_{oct}}/n_{Fe_{oct}}$$

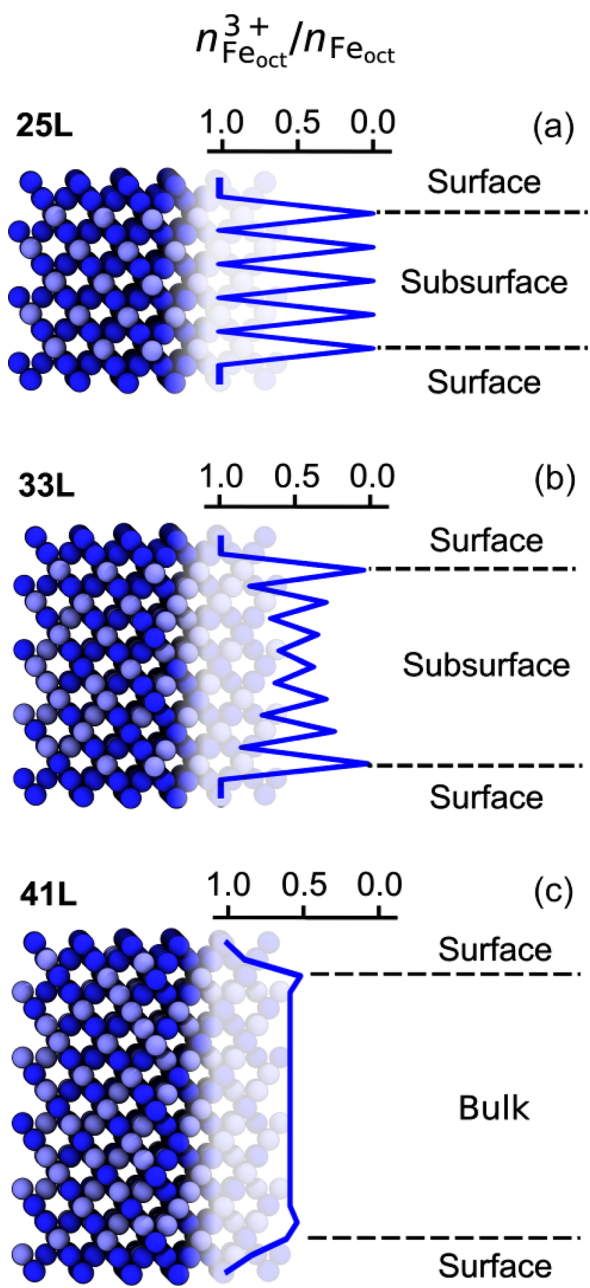

**Figure 3.** (left) Minimized oxidation-state configuration of three exemplary (001)-DBT surfaces. Number of atomic layers is indicated in the top-left corner. (right) Ratio of $Fe^{3+}_{oct}$ ions in each octahedral layer, denoted by $n^{3+}_{Fe_{oct}}/n_{Fe_{oct}}$. $Fe_{tet}$ and O are omitted. Colors: $Fe^{3+}_{oct}$, dark blue; $Fe^{2+}_{oct}$, light blue.

41L surface with a "bulk-like" middle part is energetically more favorable than the layered oxidation-state configuration of the same surface, see Figure 2b, when the atom positions are allowed to relax. This is an important finding, as it is now possible to infer how many layers are required for magnetite surfaces in general to have a bulk-like region in the center, which is an information that is difficult to access in DFT because of system-size limitations.[19,25,30,50]

We have similarly performed calculations for (001)-SCV and (111)-tet1 surfaces, which are summarized in Figures S4 and S5. Interestingly, the (001)-SCV surfaces display a bulk-like central region already for surfaces with just 13 atomic layers. For (111)-tet1 magnetite surfaces, it was previously observed that the tetrahedral iron on the surface is a $Fe^{2+}$ rather than the expected $Fe^{3+}$ due to the strong undercoordination.[22,30] These results were obtained for quite small simulation cells (i.e., two $Fe_{tet1}$ per unit cell). Our force field approach allows for much larger systems and for a minimized oxidation-state configuration, we observed that about half $Fe^{2+}$ and half $Fe^{3+}$ are present on the surface with some visible order as shown in Figure S5. Additional DFT calculations for thin, 11-layer surfaces with a 2 × 2 surface cell also resulted in mixed oxidation-state configuration for the tetrahedral irons at the surface as in the force field approach. Once again, this demonstrates the importance of larger systems in the case of magnetite and the usefulness of the oxidation-state swap approach to correctly predict the complex oxidation-state configuration found in many inverse spinel structures.

Compared to force field simulations of bulk magnetite and magnetite surfaces, nanoparticles require even more care. Apart from a potentially even greater degree of nonstoichiometry, this is due to the shape and, relatedly, the topology of nanoparticles. While enforcing the ideal $Fe_3O_4$ stoichiometry for spherical nanoparticles leads in our simulation to the formation of kinks and surfaces resembling a (001)-DBT termination once they are treated in a hybrid MC/MD approach, nanoparticles in general have edges and corners between surfaces that do not (yet) have a distinct crystallographic orientation. For the parametrization of a force field, these kinks, edges, and corners pose a major challenge in assigning partial atomic charges. Until now, these kinks, edges, or corners are often not properly accounted for, and specific iron charges are simply ignored,[51] or computationally expensive methods such as DFTB+U must be employed.[27] Here, assigning charge values becomes a simple and straightforward task, since eq 3 inherently ensures charge neutrality, while the oxidation-state configuration is minimized via oxidation swaps. In this way, any surface, edge, or corner reconstruction on nanoparticles, as shown in the example of a cubic nanoparticle, can eventually be realistically represented.

Liu and Di Valentin[27] used MD simulations on the basis of DFTB+U to study cubic and spherical nanoparticles. Cubic nanoparticles were generated by following the same approach. During annealing of the cubic nanoparticles, Liu and Di Valentin[27] observed that 4 out of 8 corners were reconstructed. In the corner reconstruction, 3 sixfold-coordinated $Fe_{oct}$ (cf. Figure 4b) were converted into 3 fourfold-coordinated $Fe_{tet}$ (cf. Figure 4c). Liu and Di Valentin[27] found that the nanoparticle reconstructed in this way was 14 meV per atom more favorable than its unreconstructed counterpart.

We first attempted to reproduce these results using our hybrid MC/MD approach. Unfortunately, it was not possible to observe the corner reconstruction using the hybrid MC/MD approach, and we manually created corresponding structures which did indeed agree well energetically with the results found in Liu and Valentin.[27] The barrier associated with this reconstruction may be overestimated by our force field approach, which effectively prevents the observation of dynamic reconstruction. Minimization of the oxidation-state configurations is achieved by using a constant $T^{MC}$ of 0.1 K, for which almost the same oxidation-state configuration is observed when compared with results from DFTB+U of Liu and Di Valentin[27] in Figure 4d. The energies of the reconstructed and unreconstructed cubic nanoparticles differ by 5.6 meV per atom, with the reconstructed cubic nanoparticle being more favorable. It is worth reiterating that this good agreement was achieved with a fairly simple force field approach.

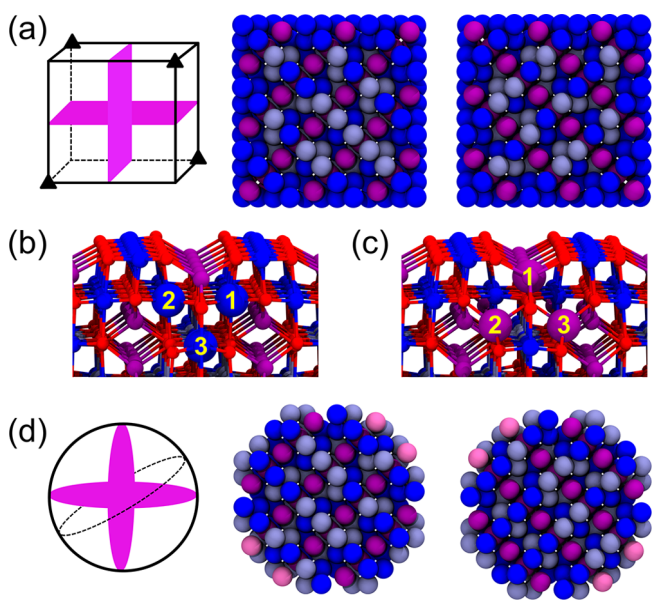

**Figure 4.** (a) Schematic representation of the corners reconstructed in a cubic nanoparticle (triangles). Pink areas correspond to cross-sectional views of a minimized oxidation-state configuration found for cubic magnetite nanoparticles on the right. (b) Corner before reconstruction; relevant $Fe_{oct}$ ions are highlighted. (c) Corner after reconstruction. (d) Cross-sectional views of a minimized oxidation-state configuration found for spherical magnetite nanoparticles. Colors: $Fe_{oct}^{3+}$ — dark blue, $Fe_{oct}^{2+}$ — light blue, $Fe_{tet}^{3+}$ — purple, $Fe_{tet}^{2+}$ — mauve, and O — red. Oxygen is omitted in (a) and (d).

In the case of a cubic nanoparticle, a distinct core–shell structure emerges where the shell layers contain only $Fe^{3+}$ ions, and the core layers have a mixture of $Fe^{2+}$ and $Fe^{3+}$ ions on octahedral iron sites. Comparing the two nanoparticles, the surface layer of the spherical nanoparticles contains more $Fe_{tet}$ ions, similar to the surfaces already discussed, but also more disorder as observed in Liu and Di Valentin.[27] Although a somewhat disordered shell structure was also observed in Liu and Di Valentin[27] for a spherical nanoparticle, the shell structure here appears to be even more disordered. This may be due to a lack of hydroxyl groups, which in Liu and Di Valentin[27] have already been added during the initial shaping to compensate for the severely undercoordinated iron on the magnetite particle. Although it is possible to include such groups in our approach, we will omit them here for simplicity and discuss them in more detail in a follow-up paper, which will also look at functionalization with organic ligands.

We have shown that the oxidation-state swap method is capable of modeling magnetite, and the results are in good agreement with electronic structure calculations, regardless of crystallographic orientation or topology and, very importantly, at much lower computational cost than reference *ab initio* calculations. When the oxidation-state swap approach presented here is restricted not only to swaps between octahedral iron sites, perfect agreement is obtained when comparing density functional theory derived oxidation-state configurations for both tetrahedral and octahedral iron sites.

Interestingly, magnetic fields have previously been observed to affect the trimeron ordering of magnetite,[52] but since our force field approach does not take into account magnetic effects, we conclude that in the temperature range studied here, magnetic effects appear to have only a negligible effect on the oxidation-state configurations. Our approach is thus able to reproduce oxidation-state configurations at temperatures around room temperature and above, which are relevant in many experiments, in reasonable agreement with DFT for many magnetite structures, from bulk to surface, and even spherical and cubic nanoparticles. In the latter case, this has been found even for unreconstructed and reconstructed corners recently found by Liu and Di Valentin.[27] However, the limitations of our approach clearly lie in the low-temperature $Cc$ structure of magnetite, which we have not been able to reproduce and which appears to be strongly influenced by magnetic effects and orbital ordering. Remarkably, such so-called trimerons found in magnetite have previously been discussed in the literature as resulting from complex quantum mechanical phenomena, i.e., Jahn–Teller distortions due to minority spin electrons in $t_{2g}$ orbitals.[29] However, such small polaron-like deformations can apparently be observed with a fairly simple force field, depending only on Coulombic interactions via point charges and van der Waals interactions represented by a Lennard-Jones potential. This suggests that the observed small polaron-like deformations may be dominated by the peculiar lattice structure of magnetite, i.e., an inverse spinel.

Another important result of our approach is that it may now be possible to determine the number of layers required for a surface to have a bulk-like region at its core. While (001)-SCV structures exhibit a bulk-like region for relatively thin surfaces, (001)-DBT counterparts require significantly larger surfaces. On the other hand, it is very interesting to note that the unit cell size plays an important role in the case of the (111)-tet surface structure. Specifically, instead of all $Fe_{tet1}$ on this surface being $Fe^{2+}$, we observed a mixture of $Fe^{2+}$ and $Fe^{3+}$ with some alternating orders of $Fe^{2+}$ and $Fe^{3+}$. Even more interesting is the change in octahedral spacing due to oxidation-state layering, which has been observed for very thin (001)-DBT magnetite surfaces and could be an interesting aspect to be verified experimentally.

The approach presented here ultimately allows the simulation of a realistic and much larger system and is therefore also suitable for studying the mechanical and structural properties of functionalized magnetite surfaces and nanoparticles and their self-assembly, which will be investigated in a follow-up work. It also opens the door to a more fundamental understanding of oxidation-state ordering in inverse spinel structures, which should not only be evident in magnetite but could also play an important role in battery materials such as lithium manganese oxide spinels, where $Mn^{2+}$ ions partially occupy the tetrahedral sites.[53−55] In conclusion, the oxidation-state swapping method, which is computationally much cheaper than any electronic structure calculations, is a promising tool for modeling magnetite and other inverse spinels with an accuracy similar to quantum chemical approaches in terms of capturing the important oxidation-state configurations.

## ■ ASSOCIATED CONTENT

### ⓢ Supporting Information

. The Supporting Information is available free of charge at https://pubs.acs.org/doi/10.1021/acs.jpclett.3c01290.

    LAMMPS example of oxidation-state swapping in bulk magnetite (ZIP)

Detailed information of the oxidation-state swap annealing procedure, Bader charges from density functional theory and magnetite force field parameters (PDF)

## AUTHOR INFORMATION

**Corresponding Author**

**Robert H. Meißner** − Institute of Polymers and Composites, Hamburg University of Technology, 21073 Hamburg, Germany; Institute of Surface Science, Helmholtz-Zentrum Hereon, 21502 Geesthacht, Germany; ◉ orcid.org/0000-0003-1926-114X; Email: robert.meissner@tuhh.de

**Authors**

**Emre Gürsoy** − Institute of Polymers and Composites, Hamburg University of Technology, 21073 Hamburg, Germany; ◉ orcid.org/0009-0008-6042-1364

**Gregor B. Vonbun-Feldbauer** − Institute of Advanced Ceramics, Hamburg University of Technology, 21073 Hamburg, Germany; ◉ orcid.org/0000-0002-9327-0450

Complete contact information is available at:
https://pubs.acs.org/10.1021/acs.jpclett.3c01290

**Notes**

The authors declare no competing financial interest.

## ACKNOWLEDGMENTS

Funded by the Deutsche Forschungsgemeinschaft (DFG, German Research Foundation) − SFB 986 − 192346071.